\documentclass[reprint,aps,prl,superscriptaddress,nofootinbib,preprintnumbers,longbibliography]{revtex4-1}

\usepackage{comment}
\usepackage{graphics}
\usepackage{bm}
\usepackage{color}
\usepackage{amsmath}
\usepackage{amssymb}
\usepackage{mathrsfs}
\usepackage{graphicx}
\usepackage[caption=false]{subfig}
\usepackage{enumitem}
\usepackage{hyperref} 

\begin{document}
\title{Quantum Logic Spectroscopy of an Electron and Positron \\for Precise Tests of the Standard Model}

\date{\today}

\author{Xing Fan}
\email{xingfan@g.harvard.edu}
\affiliation{Department of Physics, Harvard University, Cambridge, Massachusetts 02138, USA}
\affiliation{Center for Fundamental Physics, Department of Physics and Astronomy, Northwestern University, Evanston, Illinois 60208, USA}

\author{Atsushi Noguchi}
\affiliation{RIKEN Center for Quantum Computing (RQC), 2-1 Hirosawa, 351-0198, Wako, Saitama, Japan}
\affiliation{Komaba Institute for Science (KIS), The University of Tokyo, 3-8-1 Komaba, 153-8902, Meguro, Tokyo, Japan}
\affiliation{Inamori Research Institute for Science (InaRIS), 620 Suiginya, 600-8411, Kyoto, Kyoto, Japan}

\author{Kento Taniguchi}
\affiliation{RIKEN Center for Quantum Computing (RQC), 2-1 Hirosawa, 351-0198, Wako, Saitama, Japan}

\begin{abstract}
We propose a scheme for quantum logic spectroscopy of an electron or positron in a Penning trap.
An electron or positron in a spectroscopy trap is coupled to a remote logic electron or positron via a wire to achieve motional entanglement.
By separating the two traps, one can significantly improve magnetic field homogeneity, microwave characteristics, and detection sensitivity.
The proposed scheme will improve the measurement precision of the electron's and positron's magnetic moments and charge-to-mass ratios, enabling precise tests of the Standard Model of particle physics.
\end{abstract}
\maketitle
Quantum logic spectroscopy (QLS) is a technique to read out the internal state of trapped ions without a suitable level structure though coupling to a second ion\cite{QLS_PhysRevA_1990,PhysRevLett.75.4714,SpectroscopyUsingQuantumLogic,PhysRevLett.99.120502,wineland2002quantum,wineland1998experimental}.
QLS has been proposed and applied for various systems, including the state-of-the-art ion clocks\cite{PhysRevLett.98.220801,PhysRevLett.123.033201,rosenband2008frequency}, molecular ions\cite{wolf2016non,chou2017preparation,PhysRevA.109.033107}, highly charged ions\cite{micke2020coherent,king2022optical,leopold2019cryogenic}, and single protons and antiprotons\cite{cornejo2021quantum,PhysRevResearch.6.033233}.
In QLS, a logic ion is coupled to a spectroscopy ion via the motion of their charges, creating an entangled motional state\cite{QLS_PhysRevA_1990}.
The key requirement is that the charge-to-mass ratio ($q/m$) of the two ions must be sufficiently close ($\lesssim10$) to achieve efficient entanglement.
In this context, electrons and positrons are unique, as there are no ions with a close enough $q/m$.
The closest candidate would be the proton, but the ratio $(q/m)_p/(q/m)_e$ is as large as 2000, far too large to achieve effective entanglement.
Although QLS of an electron with Be$^+$ was proposed in Ref.~\cite{QLS_PhysRevA_1990}, experimental realization of the proposal would require significantly restrictive trap parameters---for example, to match their oscillation frequencies, the trap voltage must be lower than 100~mV.
Such limitations make it difficult to achieve a $g$-factor measurement exceeding the current precision of $1.3\times10^{-13}$ \cite{ElectronMagneticMoment_Fan_PRL_2023}.
To our best knowledge, QLS of an electron or a positron has not been achieved to date.

In this Letter, we propose a QLS scheme for an electron or a positron in a Penning trap using another electron or positron in a remote trap, mediated via a wire.
The key idea here is to separate the spectroscopy electron\footnote{The proposed scheme applies to both electrons and positrons; However, we use an electron as an example.} and the logic electron.
Motional entanglement created through a wire maps the quantum states between the two electrons\cite{QLS_PhysRevA_1990,osada2022feasibility,PhysRevApplied.22.024032}.
Precision measurements can be performed on the spectroscopy electron, while efficient state readout can be performed on the logic electron.
As discussed below, this scheme will reduce both the statistical and systematic uncertainties in the measurement of the electron's and positron's magnetic moments ($g$-factors) and the comparison of their charge-to-mass ratios ($q/m$).
These measurements provide the most stringent test of the Standard Model's calculations\cite{QED_C8_Lapo,QED_C8_nio,QED_C10_nio,atomsTheoryReview2019,volkov2018numerical,volkov2017new,volkov2019calculating,volkov2024calculation,kitano2024qed,odom2006new,hanneke2008new,ElectronMagneticMoment_Fan_PRL_2023}, one of the most precise determinations of the fine structure constant\cite{PhysRevLett.97.030802,morel2020determination,parker2018measurement}, the most precise test of the lepton $CPT$ symmetry\cite{bluhm1998cpt,lehnert2016cpt}, and the most precise test of the anti-gravity effects in lepton systems\cite{ulmer2015high,borchert202216}. 
The proposed QLS scheme has the potential to significantly improve all these tests.

Precise $g$-factor measurements are performed by measuring the ratio of the cyclotron frequency $\omega_c=qB/m$ and the spin frequency $\omega_s=qB/m\times(g/2)$ for an electron or positron in a Penning trap cooled to milli-Kelvin temperatures\cite{RevModPhys.58.233,ElectronMagneticMoment_Fan_PRL_2023}.
For $q/m$ comparisons, the cyclotron frequency $\omega_c$ is measured and compared between the two particles\cite{schwinberg1981trapping}.
In both cases, the electron's cyclotron motion is cooled to its quantum ground state $n_c=0$ by synchrotron radiation\cite{peil1999observing}.
Quantum jump spectroscopy of the cyclotron transition $n_c=0\rightarrow n_c=1$ is then performed to measure $\omega_c$.
Similarly, the transition between the two spin states $m_s=\pm1/2$ is used to determine $\omega_s$.

A crucial step of the spectroscopy is the readout of these transitions.
In these experiments, the axial frequency of the particle $\omega_z$ (motion parallel to the applied magnetic field) is measured via image charge detection\cite{ElectronCalorimeter}.
A quadratic magnetic field gradient $\mathbf{B}(\rho,z)=B\hat{z}+B_2\left(z^2-\rho^2/2\right)\hat{z}-B_2z\rho\hat{\rho}$ is applied, where $\hat{z}$ and $\hat{\rho}$ are the unit vectors in the cylindrical coordinate.
This field configuration causes a shift in the axial frequency (known as magnetic bottle) when the cyclotron or spin quantum number changes\cite{DehmeltMagneticBottle,fan2021driven}
\begin{equation}
\omega_z(n_c,m_s)=\omega_{z;0} + \delta\left(n_c+\frac{1}{2}+\frac{g}{2}m_s\right),\label{eq:BottleShift}
\end{equation}
where $\omega_{z;0}$ is the axial frequency when $B_2=0$, and
\begin{equation}
    \delta=\frac{\hbar e B_2}{m^2\omega_z}.\label{eq:BottleShiftDelta}
\end{equation}
This gradient, however, couples to the Boltzmann distribution of the axial motion $\langle z^2\rangle=k_BT_z/m\omega_z^2$ and induces broadening in the cyclotron and spin transitions \cite{brown1985geonium}
\begin{equation}
\Delta\omega_c=\frac{eB_2\langle z^2\rangle}{m}=\frac{eB_2}{m}\frac{k_BT_z}{m\omega_z^2}.
\label{eq:broadening}
\end{equation}
Additionally, if the radius of the magnetron motion is increased due to heating, this also couples to the $B_2$ gradient and causes further broadening\cite{PhysRevLett.122.043201}.
One could employ microwave sideband cooling to cool the electron to its axial quantum ground state $n_z=0$, but this does not evade the broadening from the heating of the magnetron motion\cite{PhysRevLett.122.043201}.
Alternatively, reducing $B_2$ to zero and detecting axial frequency shifts due to relativistic mass increase from cyclotron or spin transitions could be employed.
This ``special relativity bottle" is phenomenologically similar to Eq.~\eqref{eq:BottleShift} with
\begin{equation}
\delta_\mathrm{rel}=-\frac{\hbar\omega_c\omega_z}{2mc^2}.
\end{equation}
However, the shift from the special relativity effect is only $\delta_\mathrm{rel}=2\pi\times-0.14$~Hz  even at $B=6$~T and $\omega_z=2\pi\times 200$~MHz ($\delta_\mathrm{rel}/\omega_z\sim10^{-9}$), requiring extremely stable trap voltages and long averaging times.
In addition, to accurately correct the microwave cavity shift\cite{brown1985cyclotron,brown1985cyclotronPRL,hanneke2011cavity,fan2022_Thesis}, the Penning trap electrodes must be made cylindrical \cite{tan1989one} or spherical \cite{brown1986cyclotron} with minimal slits.
Even with the orthogonal and compensated designs\cite{gabrielse1984cylindrical}, these constraints limit the harmonicity of the electric potential, axial oscillation amplitude, and the maximum detected signal size.

\begin{figure}
    \centering
\includegraphics[width=0.8\linewidth]{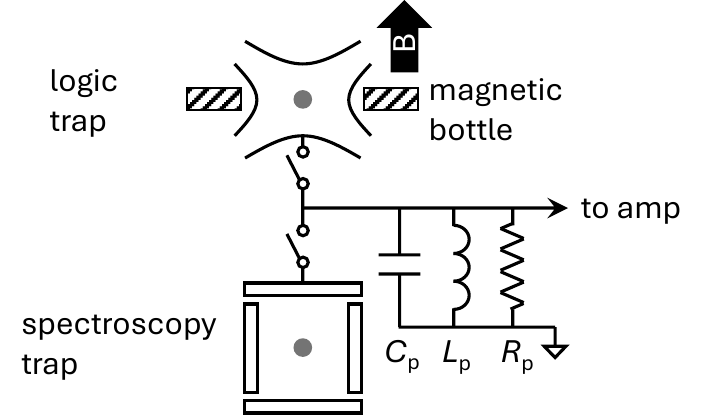} 
     \caption{Setup for the proposed QLS. A logic trap is connected to a spectroscopy trap via a wire and switches. They are also connected to an $LCR$ resonator.}
    \label{fig:Setup}
\end{figure}
To overcome these challenges, we propose a QLS scheme using two traps connected via a wire: a precision spectroscopy trap for quantum jump spectroscopy\cite{peil1999observing} and a logic trap for state readout using a motionally entangled electron (Fig.~\ref{fig:Setup}).
The spectroscopy trap is optimized for homogeneous magnetic field and microwave cavity characteristics to accurately measure $\omega_c$ and $\omega_s$.
The logic trap, on the other hand, is optimized for state readout through the magnetic bottle coupling to the axial motion.
For this purpose, a hyperbolic trap or a seven- (or more) segmented electrode trap can be used to achieve a high axial harmonicity, a much larger bottle can be used to create a large shift per quantum jump [Eq~\eqref{eq:BottleShift}], and a much smaller trap can be employed to achieve strong coupling of the axial motion to the detector\cite{ElectronCalorimeter}.
The two traps are connected by an entanglement wire that couples the axial oscillation modes of the two separate electrons\cite{daniilidis2009wiring,zurita2008wiring}.
The wire is also connected to an inductor $L_p$, forming a $LCR$ resonant circuit with the electrode capacitance $C_p$ and effective resistance $R_p$.
The $LCR$ resonator increases the coupling between the two electrons and also works as a detection circuit.
Switches---either IC-based\cite{sturm2012g}, varactor-based\cite{nagahama2016highly, volksen2022high}, or HEMT-based\cite{10.1063/5.0038005}---are installed to control the coupling of the electrons.

\begin{figure}[t]
    \centering
\includegraphics[width=0.95\linewidth]{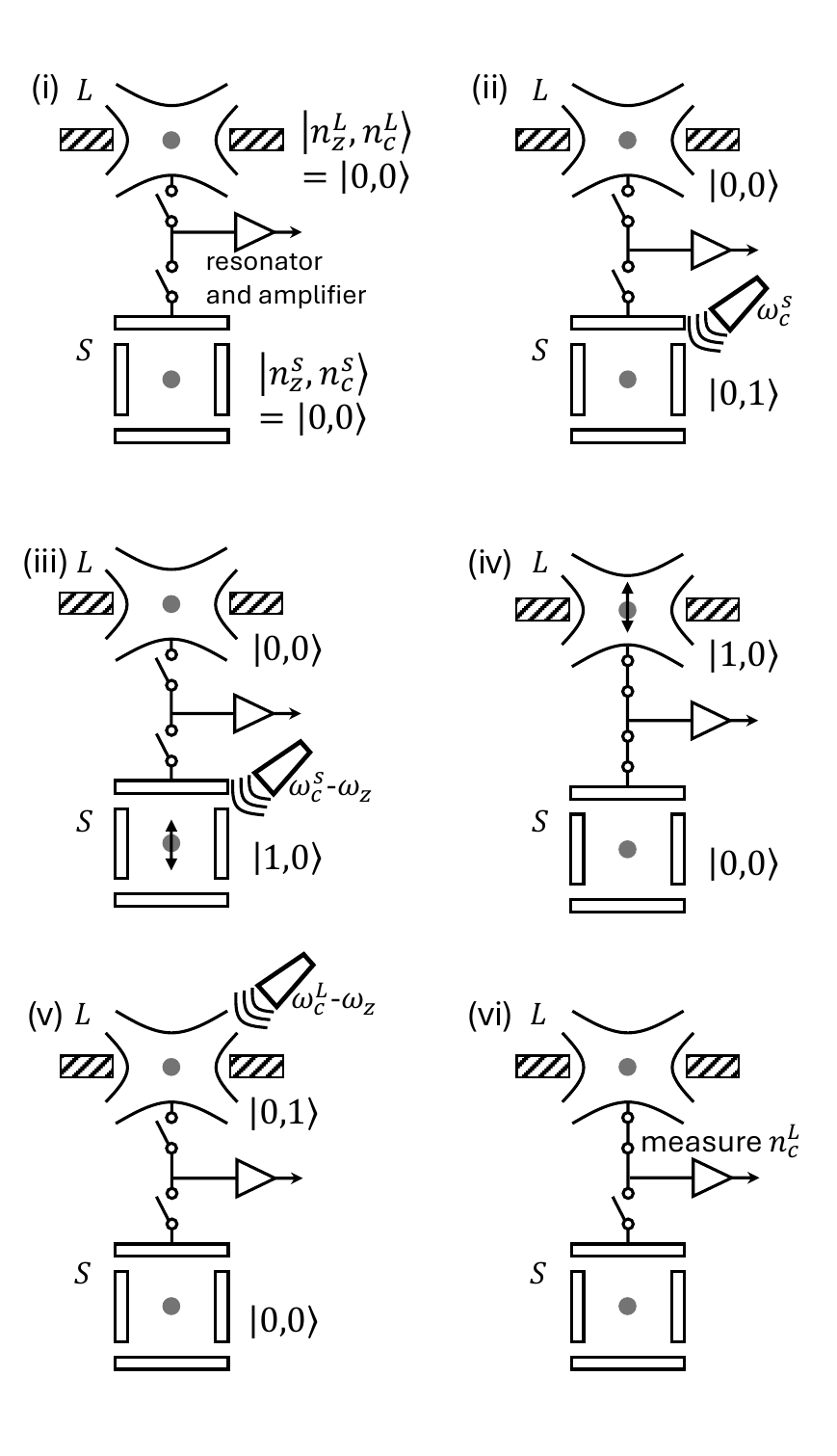} 
     \caption{Proposed QLS scheme with two separate electrons. $L$: logic trap and $S$: spectroscopy trap. See text and Ref.\cite{QLS_PhysRevA_1990} for details.}
    \label{fig:Steps}
\end{figure}
Figure.~\ref{fig:Steps} shows the measurement steps for the cyclotron transition.
To measure the spin frequency, we drive the anomaly transition $|n_c=0,m_s=+1/2\rangle\rightarrow|n_c=1,m_s=-1/2\rangle$ as in the $g$-factor measurement, which also causes a cyclotron excitation.
The spectroscopy trap and the logic trap, along with their corresponding physical quantities, are labeled $S$ and $L$, respectively.
Both electrons are tuned to the same axial frequency $\omega_z$.
The two traps will be placed in the same superconducting solenoid magnet, but the cyclotron frequencies $\omega_c^L$ and $\omega_c^S$ can be different.
The QLS sequence is performed as follows.
\begin{enumerate}[label=(\roman*)]
    \item Both electrons are decoupled from the resonator using the switches.
Sideband cooling drives $\omega_c^L-\omega_z$ and $\omega_c^S-\omega_z$ are applied to their respective traps to cool the axial motions to $n_z^\mathrm{S}=n_z^\mathrm{L}=0$.
\item A spectroscopy drive near $\omega_c^S$ is applied to the spectroscopy electron, which will excite $n_c=0\rightarrow n_c=1$, if successful.
\item A $\pi$-pulse at $\omega_c^S-\omega_z$ is applied to the spectroscopy electron to transfer $|n_z^S,n_c^S\rangle=|0, 1\rangle\rightarrow|1, 0\rangle$.
This transfer occurs only if $n_c=1$ and does not occur if $n_c=0$, as is typical in a QLS sequence.
\item The switches are turned on to couple the axial motion of the spectroscopy and the logic electrons. If $n_z^S=1$ and $n_z^L=0$, the entanglement through the wire transfers the state to $n_z^S=0$ and $n_z^L=1$ with an exchange time $t_\mathrm{ex}$.
The exchange time is calculated later.
\item Both traps are decoupled, and a $\pi$-pulse at $\omega_c^L-\omega_z$ is applied to the logic electron to transfer $|n_z^L,n_c^L\rangle=|1, 0\rangle\rightarrow|0, 1\rangle$.
Again, this occurs only if $n_c^S$ was excited in the step (ii).
\item The axial frequency of the logic electron $\omega_z^L$ is measured by coupling it to the detection resonator.
The cyclotron quantum number $n_c^L$ is determined from the bottle shift [Eq.~\eqref{eq:BottleShift}].
Since $B_2$ of the logic trap can be made large with small effect on the distant spectroscopy electron, a much faster and higher fidelity detection is possible.
\item After judging whether a transition occurred, return to step (i) and continue the spectroscopy.
\end{enumerate}

\begin{figure}[t]
    \centering
\includegraphics[width=0.65\linewidth]{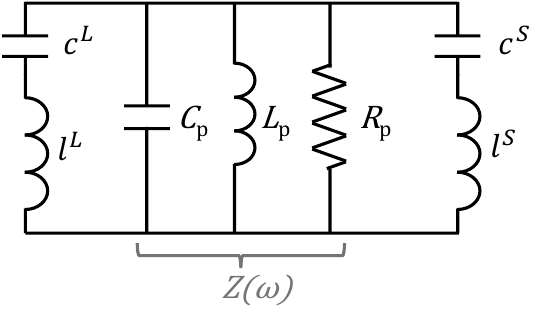} 
     \caption{Equivalent circuit model of the logic trap ($L$), spectroscopy trap ($S$), and the resonator [impedance $Z(\omega)$]\cite{ElectronCalorimeter}.}
    \label{fig:Circuit}
\end{figure}

The system can be quantitatively analyzed using an equivalent circuit model shown in Fig.~\ref{fig:Circuit}.
Here, the $LCR$ resonator between the coupling wire and the ground has a Lorentzian-shape impedance $Z(\omega)=\left[1/R_p+i\omega C_p+1/(i\omega L_p)\right]^{-1}$, with center frequency $\omega_\mathrm{res}=1/\sqrt{L_pC_p}$ and width $\Delta\omega_\mathrm{res}=1/(C_pR_p)$.
The trapped electron can be described by a series inductance and capacitance, with $l^i=m\left({2d_\mathrm{eff}^i}/e\right)^2$ and $c^i={1}/(l^i{\omega_{z;0}^i}^2)$ ($i=L$ or $S$), where $d^i_\mathrm{eff}$ is the effective trap size including the image charge parameter \cite{ElectronCalorimeter}, and $\omega_{z;0}^i$ is the original axial frequency without the resonator.
Since the resonator shifts the axial frequency by $-\frac{\mathrm{Im}\left[Z(\omega)\right]}{l^{i}}$, we set $\omega_{z;0}^i=\omega_z+\frac{\mathrm{Im}\left[Z(\omega)\right]}{l^{i}}$ by adjusting the trap voltage so that the resonator-shifted axial frequency becomes exactly $\omega_z$.

Because $\mathrm{Re}\left[Z(\omega_z)\right]\propto|\omega_z-\omega_\mathrm{res}|^{-2}$ and $\mathrm{Im}\left[Z(\omega_z)\right]\propto|\omega_z-\omega_\mathrm{res}|^{-1}$, we can detune $\omega_\mathrm{res}$ below $\omega_\mathrm{z}$ to make the circuit capacitive, $\textrm{Im}\left[Z(\omega_z)\right]\gg\textrm{Re}\left[Z(\omega_z)\right]$.
In this limit, $Z(\omega)$ can be replaced by $C_T\equiv1/(\omega_z|\textrm{Im}\left[Z(\omega_z)\right]|)$.
This circuit, as carefully analyzed in Ref.~\cite{QLS_PhysRevA_1990}, yields an exchange rate
\begin{equation}
\omega_\textrm{ex}=\frac{|\textrm{Im}\left[Z(\omega_z)\right]|}{2\sqrt{l^Ll^S}}.\label{eq:exchangeRate}
\end{equation}
A full exchange of the axial quantum numbers $n_z^i$ occurs at \begin{equation}
    t_\mathrm{ex}=\frac{\pi}{2\omega_\mathrm{ex}}.
\end{equation}

The real part of $Z(\omega_z)$ causes dissipation of the quantum state with a rate $\mathrm{Re}\left[Z(\omega_z)\right]/l^i$.
Using the maximum dissipation rate $\Gamma\equiv\mathrm{max}\left(\frac{\mathrm{Re}\left[Z(\omega_z)\right]}{l^L},\frac{\mathrm{Re}\left[Z(\omega_z)\right]}{l^S}\right)$, the requirement for achieving a reliable QLS is
\begin{equation}
    t_\mathrm{ex}\bar{n}_z\Gamma<1,
    \label{eq:QLSCondition}
\end{equation}
where $\bar{n}_z=1/\left[\exp\left(\frac{\hbar\omega_z}{k_BT}\right)-1\right]$ is the photon number density.

With typical parameters for electron Penning traps, $d^S_\mathrm{eff}=3$~mm, $d^L_\mathrm{eff}=1$~mm, $\omega_z=2\pi\times200$~MHz, $C_p=10$~pF, $R_p=500$~k$\Omega$ ($Q=6000$), and $T=10$~mK ($\bar{n}_z=0.6$)\cite{fan2022_Thesis}, by setting $\omega_\mathrm{res}=\omega_z-30\Delta\omega_\mathrm{res}$, one can achieve $t_\mathrm{ex}=160$~ms and $t_\mathrm{ex}\bar{n}_z\Gamma=0.098$.
Further suppression of heating should be possible by operating at higher axial frequencies $\omega_z$ (thus lower $\bar{n}_z$) or by improving the resonator's $R_p$ (or equivalently $Q$).

The anomalous heating of the electron's axial motion in a Penning trap could reduce the coherence.
However, due to the absence of RF fields in Penning traps, the high axial frequency, the large millimeter-scale ion-electrode distance, and the cryogenic environment, we expect the heating rates to be much lower than those of atomic ions in Paul traps, as demonstrated in Ref.~\cite{PhysRevLett.122.043201}.
Scaling from typical electric field noise \cite{RevModPhys.87.1419} with a conservative estimate of $S_E=10^{-12}~\mathrm{V}^2\mathrm{m}^{-2}\mathrm{Hz}^{-1}\times\Big(\frac{\omega/2\pi}{1~\mathrm{MHz}}\Big)^{-1}\Big(\frac{d}{100~\mu\mathrm{m}}\Big)^{-2}\Big(\frac{T}{6~\mathrm{K}}\Big)^{0.5}$, the heating rate in the electron's Penning trap will be less than $1~\mathrm{quanta}/s$.
Thus, we believe that the anomalous heating is not dominant.

\begin{figure}
    \centering
\includegraphics[width=0.9\linewidth]{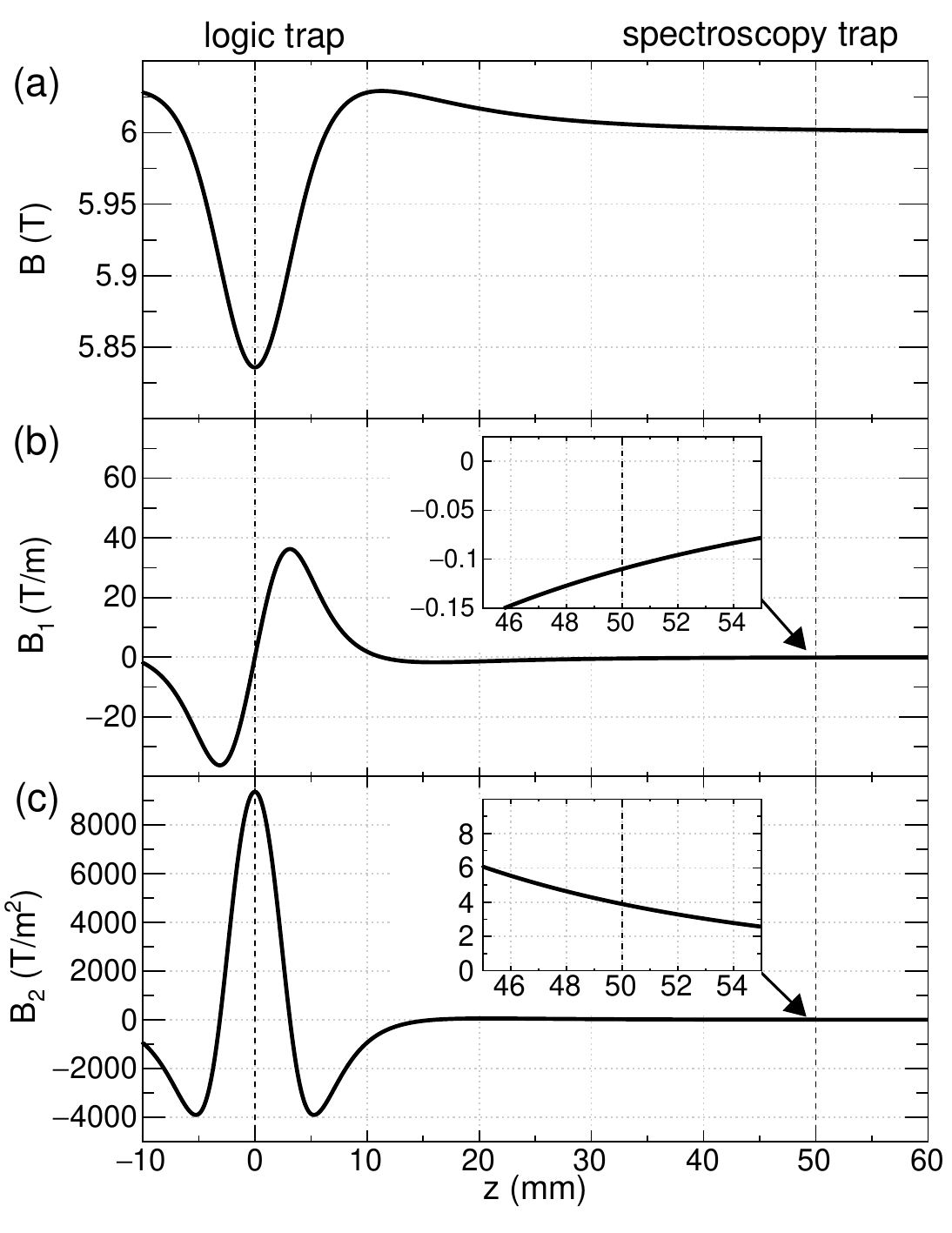} 
     \caption{Magnetic field $B$, its first gradient $B_1=\frac{dB(z)}{dz}$, and second gradient $B_2=\frac{1}{2}\frac{d^2B(z)}{dz^2}$ as a function of the distance from the logic trap. A cobalt-iron bottle with susceptibility $\chi=20,000$, inner radius 5~mm, outer radius 15~mm, and height~5~mm is placed at $z=0$. The location of the logic trap and the spectroscopy trap are indicated by the dotted lines. The quadratic gradient at the spectroscopy trap $B_2=4$~T/m$^2$ is 75 times smaller than the 2023 $g$-factor measurement \cite{ElectronMagneticMoment_Fan_PRL_2023}.}
    \label{fig:BField}
\end{figure}

The advantage of the proposed QLS scheme for the $g$-factor measurement comes from several factors.
As shown in Eq~\eqref{eq:broadening}, the linewidth of the cyclotron transition is determined by $B_2$.
In the QLS scheme, the magnetic bottle can be placed at the logic trap, far from the spectroscopy trap, for instance, by $l=5$~cm.
At $B=6$~T, a cobalt-iron ring with susceptibility $\chi=20,000$, inner radius 5~mm, outer radius 15~mm, and height 5~mm, generates $B_2=9,000$~T/m$^2$ (Fig.~\ref{fig:BField}), 30 times larger than in the 2023 $g$-factor measurement\cite{ElectronMagneticMoment_Fan_PRL_2023}.
The corresponding bottle shift, $\delta=2\pi\times 23$~Hz [Eq.~\eqref{eq:BottleShiftDelta}], allows about 20 times faster detection and less stringent requirements on the voltage stability.
In contrast, the $B_2$ gradient at the spectroscopy trap will be as small as $B_2=4$~T/m$^2$, 75 times smaller than the gradient in the 2023 $g$-factor measurement\cite{ElectronMagneticMoment_Fan_PRL_2023}, and could be easily further reduced by shim coils.
With these improvements, the linewidth will be limited by the fluctuation of the superconducting magnet's field ($\Delta B/B\sim10^{-10}$ on a minute timescale\cite{PhysRevLett.124.113001,fan2019gaseous,gabrielse1988self}), and we anticipate a statistical uncertainty of $\delta g/g\sim10^{-14}$ in one day of measurement.

One of the largest systematic errors in the $g$-factor measurement is the microwave cavity shift\cite{brown1985cyclotronPRL,hanneke2011cavity,gabrielse1985observation} from the coupling between the electron's cyclotron motion and the Penning trap's microwave resonances (the so-called Lamb shift in cavity quantum electrodynamics).
In the traditional one-trap scheme, to achieve a good electric potential harmonicity and good microwave characteristics, a cylindrical trap with four horizontal slits was used\cite{gabrielse1984cylindrical}.
The slits, however, causes distortion of the microwave cavity modes and generates the systematic error\cite{hanneke2011cavity}.
In the QLS scheme, the harmonicity of the trap potential and the microwave characteristics can be optimized independently for the logic trap and the spectroscopy trap, respectively.
For the logic trap, a seven-segmented cylindrical traps\cite{alphatrap} or a hyperbolic trap\cite{PhysRevA.27.2277} can be used to achieve a higher potential harmonicity.
A much smaller trap with $d_\mathrm{eff}^L=1$~mm will also increase the axial detection efficiency by about a factor of 10.
In contrast, the spectroscopy trap does not require a highly harmonic potential and can instead be optimized for microwave cavity characteristics.
For example, a spherical trap with two slits\cite{brown1986cyclotron} can be employed.
The spherical geometry will reduce the number of coupling microwave modes by a factor of 4, and the two-slit design will reduce mode distortion by a factor of 2.
We anticipate to improve the dominant systematic error from microwave cavity to $\delta g/g<3\times10^{-14}$.

The proposed scheme can also be used for a comparison of the charge-to-mass ratio of an electron and a positron.
In this comparison, the cyclotron frequency $\omega_c=qB/m$ of an electron and a positron is measured and compared in the same trap.
The largest systematic error comes from temporal variation of the magnetic field\cite{van1999ultrastable} and the position displacement of an electron and a positron by non-reversing trap potential, coupled to the magnetic field gradients $B_1$ and $B_2$\cite{schwinberg1981trapping,ulmer2015high,PhysRevLett.82.3198}.
The temporal variation can be suppressed by the fast readout in the QLS scheme, as discussed above.
The shift from displacement can also be suppressed due to the highly homogeneous magnetic field in the spectroscopy trap.
One could also move an electron or positron in the spectroscopy trap as a magnetic field sensor and optimize the homogeneity further.

In principle, the QLS scheme could be applied for proton and antiproton systems as well\cite{gabrielse1999precision,gabrielse1990thousandfold,gabrielse1989cooling,gabrielse1995special,smorra2017parts,schneider2017double,smorra2015base,latacz2023base,ulmer2011Thesis}.
In Eq.~\eqref{eq:exchangeRate}, $l^i$ will be about 2000 times larger due to the heavier mass of (anti-)protons, but the impedance $Z(\omega_z)$ will also be significantly higher by using the superconducting resonators\cite{nagahama2016highly,ulmer2009quality}.
These two factors nearly cancel out, resulting in approximately the same exchange rate. 
However, the lower axial frequency ($\omega_z\simeq 2\pi\times1$~MHz) leads to a high photon number ($\bar{n}_z=300$) even at 10~mK, making the requirement for dissipation difficult [Eq.~\eqref{eq:QLSCondition}].
Indeed, one can use Be$^+$ ions as the logic ion for (anti-)protons which have a charge-to-mass ratio of 9 and a closed optical transition\cite{cornejo2021quantum,bohman2021sympathetic}.
Therefore, we think that the benefit of the wire-mediated QLS scheme in the (anti-)proton systems is not as significant as in the electron and positron systems.

In conclusion, we propose a wire-mediated quantum logic spectroscopy of an electron or positron using another electron or positron in a remote logic trap.
By separating the functions of the two traps, the magnetic field homogeneity, detection sensitivity, and microwave characteristics can be significantly improved.
We believe that this scheme will substantially reduce the uncertainties in the $g$-factor measurements and the charge-to-mass ratio comparisons for electrons and positrons.
The resulting improvements will determine the fine structure constant, test the Standard Model's calculations, and probe the Lepton $CPT$ asymmetry with unprecedented precision.

\begin{acknowledgments}
We thank G.~Gabrielse, B.~A.~D.~Sukra, and S.~Moroch for useful discussions. This work is supported by NSF Grants No.~PHY-2110565 and No.~PHY-2409434 and by the Masason Foundation.
\end{acknowledgments}

\bibliography{PenningTrapExperimentRefs}

\end{document}